\documentclass[sn-mathphys,referee]{sn-jnl}
\usepackage{xcolor}
\usepackage[numbers]{natbib}
\usepackage[inline]{enumitem}

\jyear{2022}%
\begin{document}

\title[Meaningful human control: actionable properties for AI system development]{Meaningful human control: actionable properties for AI system development}

\author*[1,2]{Luciano {Cavalcante Siebert}}
\email{L.CavalcanteSiebert@tudelft.nl}
\equalcont{Co-first authors.}
\author[1,3,]{Maria$\,$Luce Lupetti}
\equalcont{Co-first authors.}

\author[1,2]{Evgeni Aizenberg}
\equalcont{Co-first authors.}

\author[1,4]{Niek Beckers}
\equalcont{Co-first authors.}

\author[1,4]{Arkady Zgonnikov}
\equalcont{Co-first authors.}
\author[1,5]{Herman Veluwenkamp}
\author[1,4]{David Abbink}
\author[1,3]{Elisa Giaccardi}
\author[1,2]{Geert-\hskip1pt Jan Houben}
\author[1,2]{Catholijn M. Jonker}
\author[1,5]{Jeroen {van den Hoven}}
\author[1]{Deborah Forster}
\author[1,2]{Reginald L. Lagendijk}

\affil[1]{\orgdiv{AiTech Interdisciplinary Research Program on Meaningful Human Control},
    \orgname{Delft University of Technology},
    \city{Delft},
    \country{The Netherlands}}
\affil[2]{\orgdiv{Faculty of Electrical Engineering, Mathematics and Computer Science},
    \orgname{Delft University of Technology},
    \city{Delft},
    \country{The Netherlands}}    
\affil[3]{\orgdiv{Faculty of Industrial Design Engineering},
    \orgname{Delft University of Technology},
    \city{Delft},
    \country{The Netherlands}}   
\affil[4]{\orgdiv{Faculty of Mechanical, Maritime and Materials Engineering},
    \orgname{Delft University of Technology},
    \city{Delft},
    \country{The Netherlands}} 
\affil[5]{\orgdiv{Faculty of Technology, Policy and Management},
    \orgname{Delft University of Technology},
    \city{Delft},
    \country{The Netherlands}}

\abstract{How can humans remain in control of artificial intelligence (AI)-based systems designed to perform tasks autonomously? Such systems are increasingly ubiquitous, creating benefits - but also undesirable situations where moral responsibility for their actions cannot be properly attributed to any particular person or group. The concept of meaningful human control has been proposed to address responsibility gaps and mitigate them by establishing conditions that enable a proper attribution of responsibility for humans; however, clear requirements for researchers, designers, and engineers are yet inexistent, making the development of AI-based systems that remain under meaningful human control challenging. In this paper, we address the gap between philosophical theory and engineering practice by identifying, through an iterative process of abductive thinking, four actionable properties for AI-based systems under meaningful human control, which we discuss making use of two applications scenarios: automated vehicles and AI-based hiring. First, a system in which humans and AI algorithms interact should have an explicitly defined domain of morally loaded situations within which the system ought to operate. Second, humans and AI agents within the system should have appropriate and mutually compatible representations. Third, responsibility attributed to a human should be commensurate with that human’s ability and authority to control the system. Fourth, there should be explicit links between the actions of the AI agents and actions of humans who are aware of their moral responsibility. We argue that these four properties will support practically-minded professionals to take concrete steps toward designing and engineering for AI systems that facilitate meaningful human control.}


\maketitle

\section{Introduction}
\label{Introduction}

Deploying AI algorithms in human-inhabited environments comes with the risk of inappropriate, undesirable, or unpredictable consequences \cite{floridi_how_2020,stinson2022algorithms}. The misinterpreted skills of AI systems, combined with their rapid impact in public and private spheres of life, can lead to situations with a clear misalignment between human moral values and societal norms \cite{coeckelbergh_ai_2020,cruz_shared_2019,jobin_global_2019,umbrello_valuesensitive_2018}, and where moral responsibility for such undesired impacts can often not be properly attributed to any person \cite{matthias_responsibility_2004}.


How can designers, users, or other human agents be morally responsible for systems 

that are designed to perform tasks, learn, and adapt without direct human control? The strong technical drive towards achieving systems which can act independently from human control in more scenarios, does not necessarily include considerations about the socio-technical consequences of implementing these systems, especially in terms of facilitating moral responsibility. In agreement with the hybrid intelligence community \cite{akata_research_2020}, we believe a stronger focus on human-AI systems (systems in which humans and AI algorithms interact during operation) is needed to address the complex design issue of ethical use and implementation of AI~\cite{dignum_ethics_2018}. The very features that make AI algorithms useful complicate their assessment and predictability in the complex socio-technical context in which they operate - which changes over time. As a result, all systems based on AI, especially those with so-called higher ``levels of autonomy''~\footnote{``Level of autonomy'' is a complex construct. In line with Bradshaw's seven deadly myths of autonomy \cite{bradshaw2013seven}, we acknowledge that measuring autonomy on a single ordered scale of increasing levels is insufficient because it lacks context, is not human-centred, and disregard functional differences, among other reasons.}, 
can and should be designed for appropriate human responsibility  \cite{santonidesio_meaningful_2018}. The holy grail is to design these systems in a manner that can mitigate the occurrence of situations that the manufacturer was in principle unable to anticipate, and that users were not able to appropriately influence or even realize. 

The problem of designing for human responsibility over human-AI systems is challenging because such systems operate in complex social infrastructures that include organizational processes with both human-to-human and human-AI interactions, policy, and law. Designing for moral responsibility therefore requires a systemic, socio-technical perspective that jointly considers the interaction between all these elements \cite{mecacci_meaningful_2020}. This fundamental challenge of intertwined social, physical, and technical infrastructures does not exclusively concern AI: societies have settled on morally acceptable solutions for ubiquitous technology in other domains, such as medicine and aviation safety. 

However, these solutions do not readily generalize to systems based on AI algorithms, due to properties such as:   
\begin{enumerate*}[label=(\arabic*)]
    \item learning abilities; 
    \item black-box nature; 
    \item impact on many stakeholders (even those not using the systems themselves); and 
    \item 
    
    autonomous or semi-autonomous decision-making features. 
    
\end{enumerate*}
First, AI agents can demonstrate novel behavior through learning from historical data and continuous learning via interactions with the world and other agents. Because the world we are concerned with is an open system with respect to the the agents’ perceptions and actions, the behavior of human-AI systems cannot be predicted with precision over time \cite{johnson_abrupt_2013,rahwan_machine_2019}. 
Second, the agent’s decision-making process may be difficult to explain and predict, even for its programmer \cite{europeanparliamentaryresearchserviceeprs_ethics_2020}, complicating responsibility attribution for its consequences. Third, as AI agents may interact with multiple users, which have different levels of expertise, different preferences, and understanding, responsibility can become a diffuse concept for which no one feels morally engaged. This may be further exacerbated when AI agent’s autonomous features are overestimated by those interacting with it. As the system's design process may overlap with implementation and use \cite{giaccardi_technology_2020}, interactions may end up including humans who did not choose to be involved in its use, as in the case of sidewalk pedestrians interacting with automated vehicles. Fourth, as systems based on AI with increasing autonomous decision-making features operate with reduced or even no meaningful supervision, undesirable impacts might be perceived only in hindsight. Learning abilities, opacity, interaction with many stakeholders, and autonomous or semi-autonomous features are just four of the prominent issues, which emerge as algorithms interact with social environments.

To design for moral responsibility and human control is particularly important as quick development and immediate deployment ``in the wild'' \cite{chen_observing_2017}, instead of regulated tests procedures, is urging academia and governments to take a stance in defining visions for trustworthy AI \cite{jobin_global_2019}. In fact, even if the ``move fast and break things'' mantra was considered acceptable and received wide consensus for driving digital innovation in the last decade, the same cannot be for AI with autonomous features \cite{taplin_move_2017}. A failure of an AI agent is not a ``404 error page''. It is a car accident, most likely with fatalities \cite{johnston_boeing_2019,serter_foreseeable_2017}; it is an unfair and discriminatory distribution of wealth and services \cite{korinek_artificial_2017}; it is an unjust crime accusation based on ethnicity \cite{angwin_machine_2016,sweeney_discrimination_2013}. Designers and developers of AI systems can only tackle this challenge by acknowledging upfront that successful attribution and apportioning of responsibility is not a matter of fortuitous allocation of praise or blame.


The concept of \textit{meaningful human control} \cite{article36_key_2014,article36_killing_2015,horowitz_meaningful_2015,santonidesio_meaningful_2018} was first proposed to address the problem of responsibility gaps in autonomous weapon systems, but is becoming a central concept when discussing responsible AI \cite{santonidesio_meaningful_2018}.The core idea is that humans should ultimately remain in control of, and thus morally responsible for, the behavior of human-AI systems
\footnote{Meaningful human control relates not only to the engineering of the AI agent, but also to the design of the socio-technical environment that surrounds it, including social and institutional practices \cite{behymer_autonomous_2016,santonidesio_meaningful_2018,santonidesio_four_2021}. As \cite{mecacci_meaningful_2020} elaborate, ``[intelligent] devices themselves play an important role but cannot be considered without accounting for the numerous human agents, their physical environment, and the social, political and legal infrastructures in which they are embedded.''}.
Nevertheless, meaningful human control has also received the critique to be an ill-defined concept \cite{cummings_lethal_2019} that ignores operational context \cite{ekelhof_moving_2019} and does not provide concrete design guidelines \cite{mecacci_meaningful_2020}. 

This article aims to contribute to closing the gap between the theory of meaningful human control, as proposed by \cite{santonidesio_meaningful_2018}, and the practice of designing and developing human-AI systems by proposing four actionable properties that can be addressed throughout the system's lifecycle. We start by unpacking the philosophical concept of meaningful human control (Section~\ref{sec_related}). We then present a set of four properties that were generated through an iterative process of abductive thinking that combined the different disciplinary perspectives of the authors (engineering, computer science, philosophy of technology, ethnography and design). We describe each property and illustrate how each of them helps defining whether and to what extent a human-AI system is under meaningful human control. We also suggest concrete methods and tools that can support addressing each property and illustrate them with respect to two case studies: automated vehicles and AI-based hiring (Section~\ref{sec_properties}).
Finally, we discuss the systemic and socio-technical nature of these properties and the need for transdisciplinary practices (Section~\ref{sec_disc}) and conclude the paper (Section~\ref{sec_concl}).

\section{Meaningful human control: tracking and tracing}\label{sec_related}
The concept of meaningful human control was coined in the debates on autonomous weapon systems  \cite{article36_key_2014,article36_killing_2015}. At the heart of this concept is the idea that humans need to retain control and moral responsibility over autonomous systems. This discussion is no longer exclusive to the military domain. Meaningful human control is increasingly relevant in other domains as AI agents become more ubiquitous and autonomous, especially in non-forgiving scenarios in which fundamental human rights and safety are at stake. The concept has already been discussed on the context of automated vehicles \cite{beckers_intelligent_2019,calvert_human_2019,mecacci_meaningful_2020}, including truck platooning \cite{calvert_full_2018}, surgical robots \cite{ficuciello_autonomy_2019}, smart home systems \cite{umbrello_meaningful_2020}, medical diagnosis \cite{braun_primer_2020}, and content moderation in social media \cite{wagner_liable_2019}.

Although many authors agree on the need for some form of human control over AI agents  \cite{article36_key_2014,article36_killing_2015,ekelhof_moving_2019, santonidesio_meaningful_2018}, these same authors may diverge and often disagree about what makes human control meaningful. 
Observing the theoretical challenges of specifying what meaningful human control means, Santoni de Sio and Van den Hoven \cite{santonidesio_meaningful_2018} laid out a foundation for a theory of meaningful human control with an adaptation of Fischer and Ravizza's \cite{fischer_responsibility_1998} philosophical account on guidance control, moral responsibility, and free will. Following the ideals of responsible innovation \cite{vandenhoven_value_2013} and value-sensitive design \cite{friedman_value_2019}, a centerpiece of Santoni de Sio's and Van den Hoven's conception of meaningful human control is two necessary conditions for meaningful human control, tracking and tracing:


\begin{enumerate}[label=(\arabic*)]
    \item \textit{Tracking} condition: in order to be under meaningful human control, a human-AI system should be responsive to the human moral reasons relevant in the circumstances. A human-AI system that fulfills this condition is said to track the relevant human moral reasons.
    \item \textit{Tracing} condition: in order for a human-AI system to be under meaningful human control, its behavior, capabilities, and possible effects in the world should be traceable to a proper moral and technical understanding on the part of at least one relevant human agent who designs or interacts with the system.
\end{enumerate}

In the tracking condition, control is said to be meaningful when the system's performance co-varies with the reasons of the relevant person or persons, like a mercury column in a thermometer co-varies with the temperature in the room. When air humidity varies, but the temperature remains constant, we expect no change in the mercury column, since it only tracks temperature. Similarly, when someone always accepts a new job only because the salary is higher, that person tracks financial gain, not necessarily the job’s intrinsic reward. 

For the tracing condition, it is required that this relevant person or persons are in a position to have a proper moral and technical understanding of the system. That would not be the case if the thermometer would randomly induce changes in the mercury column, or if the concept of the mercury expanding or contracting is entirely unknown to the person. Similarly, a supervisor of an automated vehicle that does not understand traffic rules, would not have such understanding.

The tracking and tracing conditions take the concept of meaningful human control one step closer to support practical design and development because they provide high-level design requirements for a human-AI system to be under meaningful human control. 
Building on this conception, researchers developed frameworks to analyze and quantify factors affecting meaningful human control for automated vehicles~\cite{heikoop_human_2019,calvert_conceptual_2020,mecacci_meaningful_2020}. 
However, a description of general system-level properties that could support operationalization of tracking and tracing conditions in diverse contexts is yet to be specified.

\section{Four properties of human-AI systems under meaningful human control}\label{sec_properties}


The tracking and tracing conditions \cite{santonidesio_meaningful_2018} provide a philosophical grounding for informing the development of human-AI systems under meaningful human control. Yet, translating these philosophical concepts into a concrete design and engineering practice is far from trivial. For instance, the tracking condition suggests that a human-AI system should be responsive to the moral reasons of a relevant human. But, \textit{how do we define the relevant human in a given circumstance? How should a given AI system recognize a moral reasoning? Does the condition imply that every AI system should be designed to be morally sensitive \cite{wallach2020machine}?} The tracing condition implies the necessity of a proper moral and technical understanding from at least one relevant human interacting and designing the system. \textit{Does this imply that the AI system should be able to recognize if and when an interacting human has such proper moral and technical understanding?} Or, \textit{does this imply that we need protocols for the design and use of AI systems that define if and when a human can and must have such understanding?}

\begin{figure}[!ht]
    \centering
    \includegraphics[width=\textwidth]{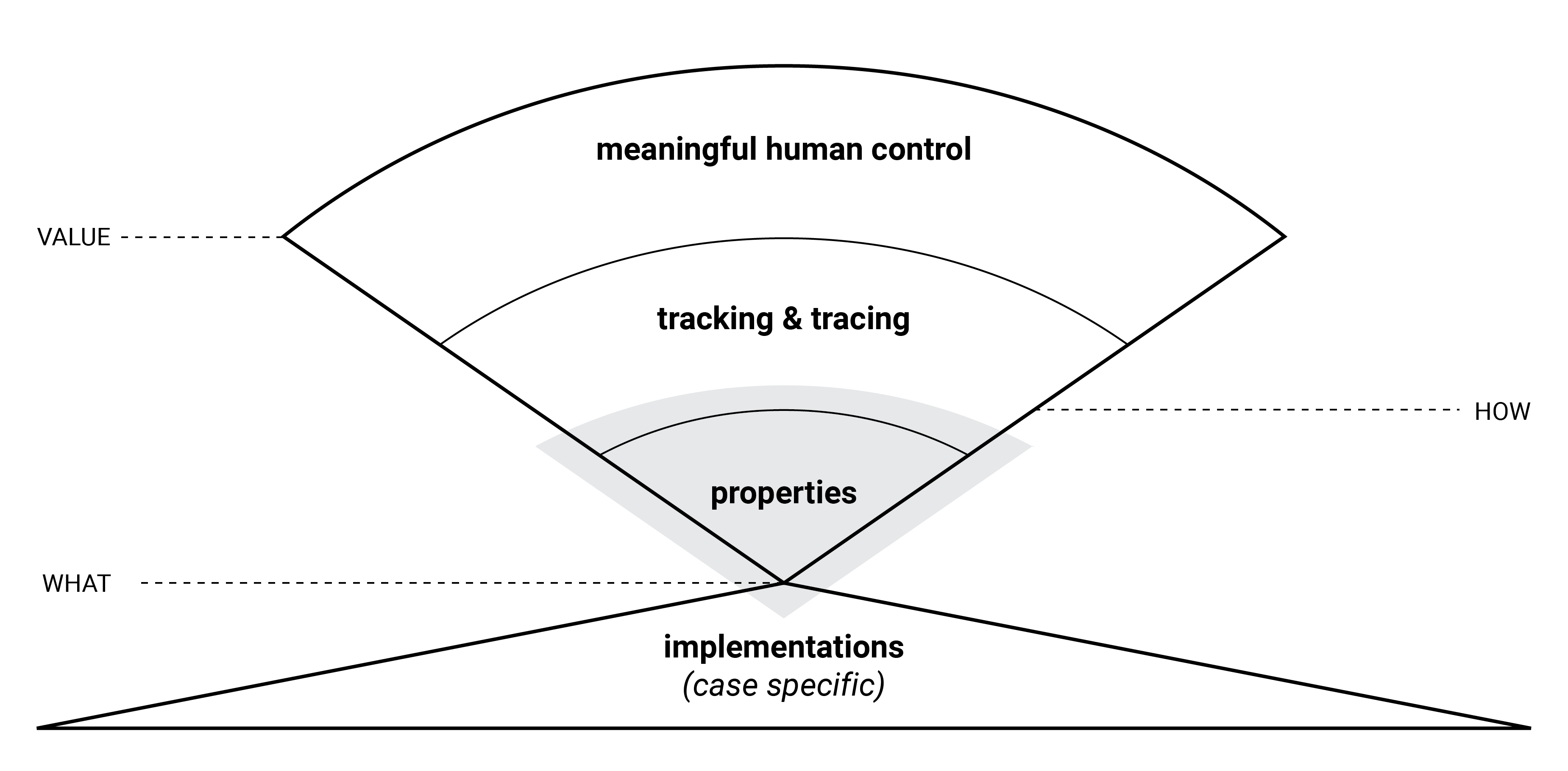}
    \caption{The diagram, based on the framework of abductive thinking by Dorst \cite{dorst2010nature}, illustrating the positioning of the abductive thinking we performed in search for a solution space that would meet the claimed need for meaningful human control.}
    \label{fig:whathowvalue}
\end{figure}

In an effort to answer these questions --- and more --- the authors, a group of researchers from various backgrounds (engineering, computer science, philosophy of technology, ethnography and design), engaged in an iterative process of abductive thinking \cite{timmermans2012theory}. Specifically, we built on Dorst's conceptual framework of abductive thinking \cite{dorst2010nature}, where both a desired value (meaningful human control) and a working principle (tracking \& tracing conditions) are known, to brainstorm ideas of what the solutions to achieve these might be. The generated ideas were then grouped into thematic areas and synthesized into actionable properties. It has to be noted, however, that although this work explores the solution space of the framework, our aim is rather to provide a contribution that sits on a meta-level, in between the what and the how (see Figure \ref{fig:whathowvalue}).

Specifically, as a result of our abductive thinking where we collectively reflected on what strategic and engineering solutions would enable the two necessary conditions of meaningful human control \cite{santonidesio_meaningful_2018}, we identified a set of four actionable properties.
In the following subsections we describe these properties in detail and illustrate their practical implications. To better highlight the properties and the implications, we make use of two example application scenarios: automated vehicles
\footnote{\textit{Automated vehicle scenario:} A conditionally automated vehicle that can perform operational aspects of the driving task (e.g., lane keeping or adaptive cruise control) as well as tactical aspects: detecting events and objects on the road and responding to them, and interacting with human pedestrians and other vehicles. Under normal circumstances, the automated vehicle can complete a whole trip without interventions from the human driver; the manufacturer emphasizes these autonomous features in their marketing and promotional materials. The driver, however, is required to constantly supervise the system. There is no requirement for the driver to keep their hands on the steering wheel, but the driver must remain alert at all times and be able to take over operational control at the request of the automation system. The vehicle does not actively monitor the driver state, but in case a human intervention is required, it attracts the driver's attention through a visual alert message and a loud auditory signal. The automated driving system used in the vehicle relies on machine learning-based object recognition and behavior prediction components, which were trained on the data obtained during extensive testing on public roads.}
and AI-based hiring
\footnote{\textit{AI-based hiring scenario:} Job candidates applying for a vacancy go through an automated video interview where they record their answers to questions formulated ahead of time by the employer. After the interview is completed, an AI agent applies machine learning methods to quantify candidates’ suitability for the job by correlating their facial expressions, choice of words, and voice tone to personal traits such as creativity, willingness to learn, and conscientiousness. To tailor the AI agent towards the context-specific preferences of the employer, the machine learning algorithms were trained on video interviews performed with current employees and their respective annual performance evaluations. The employer sets a threshold for a passing score, and based on the scores outputted by the AI agent, a list of candidates who pass to the next selection round is automatically compiled. The candidates do not see the score they were assigned. Neither the candidate, nor the employer, receive an explanation of how the scores were computed. The employer considers the human-AI system to be a cost-effective solution for what has previously been a time-consuming first-round selection process that required hiring additional screening staff. In addition, the employer seeks to increase diversity at the company and considers AI-based selection to be less prone to discriminatory biases.}
. Both cases manifest an urgent need for meaningful human control in non-forgiving scenarios that strongly impact people’s lives (e.g., bodily harm, unfair decisions, discrimination), and their differences with respect to time constraints, embodiment, and involved stakeholders juxtapose different aspects of realizing these properties in human-AI systems.

\noindent

\subsection{Property 1. The human-AI system has an explicit moral operational design domain (moral ODD) and the AI agent adheres to the boundaries of this domain}

As the human-AI system has to be ``responsive to relevant human moral reasons'' (i.e., the tracking condition), we need to identify the relevant humans, their relevant (moral) reasons, and the circumstances in which these reasons are relevant. To this end, specifying the technical conditions in which the system is designed to operate is not sufficient. Designers should consider a larger design space, one that captures also the values and societal norms that must be considered and respected during both design and operation. 

Building on the concept of operational design domain (ODD) which originates in the automotive domain~\cite{sae_taxonomy_2018}, we name this larger design space the \textit{moral operational design domain} (moral ODD). The concept of ODD is often used in the context of automated driving and refers to a set of contextual conditions under which a driving automation system is designed to function: outside of it a human driver is responsible. 
Specific contextual properties of the automotive ODD typically include factors like road structure, road users, road obstacles and environmental conditions (material elements), as well as human-vehicle interactions and expected vehicle interactions with pedestrians (relational elements)  \cite{czarnecki_operational_2018}. In terms of legal responsibility, the ODD constitutes a selection of operation scenarios that can be safely managed \cite{koopman_how_2019} by the automation and in which undesired consequences are minimized \cite{czarnecki_operational_2018}. As such, we believe it is a valuable concept to extend beyond automated vehicles, but for human-AI systems in general.

\begin{figure}[!ht]
    \centering
    \includegraphics[width=\textwidth]{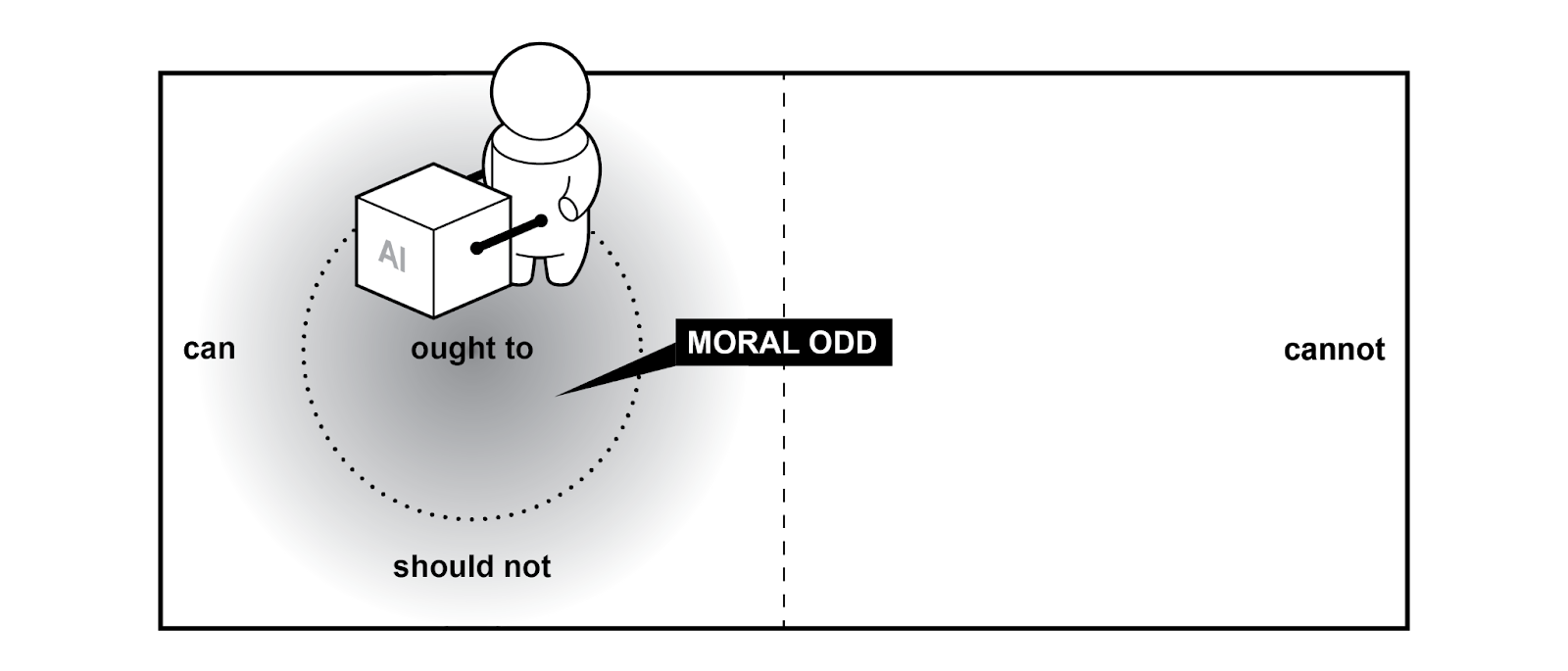}
    \caption{Property 1: Moral ODD. The human-AI system should operate within the boundaries of what it can do (for both the human and the AI agent) and within the moral boundaries of what it ought to do, i.e. the human-AI system should act according to the relevant moral reasons of the relevant stakeholders.}
    \label{fig:prop1}
\end{figure}

The current conceptualization of the ODD strongly focuses on the technical aspects of operation and the goal to extend the context boundaries of the ODD.  However, consideration of the wider societal implications is lacking. Similar to Burton et al. \cite{burton_mind_2020}, we argue that the concept of ODD should also emphasize the broader social and ethical implications. We propose this extended concept of ODD so that functional considerations of where and when a human-AI system \textit{can} operate, are seconded and complemented to the definition of the domain in which a system is \textit{ought} or \textit{should not} operate from a moral perspective (Figure \ref{fig:prop1}).

A simple example of a hammer illustrates the difference between the ``\textit{can}'' (e.g., material and relational elements)  and the ``\textit{ought to}'' dimensions (e.g., moral elements). From a purely functional perspective, a hammer ``can'' be used as a weapon against another person. However, the morally acceptable use of a hammer is for hammering nails (ought to), not to injure other people (should not). Common sense already tells us that the use of a hammer as a weapon is in most cases morally unacceptable (can but should not). It is clear that the responsibility for proper use lies with the user, not the manufacturer (except in cases where the hammer clearly does not function properly, e.g., the head suddenly comes loose from the handle and injures a person).

In a scenario involving complex human-AI systems, this is often much less clear cut. In the automated vehicle case, the moral ODD could contain moral reasons representing safety (e.g., avoid road accidents), efficiency (e.g., reduce travel time), and personal freedom (e.g., enhance independence for seniors), to name just a few. In the AI-based hiring context, moral reasons could include, from the employer's side, reducing discrimination or increasing the number of applicants in the recruitment process, while for the applicants' side autonomy over self-representation could be considered very relevant. In both contexts, however, there might be tensions among different moral reasons and stakeholders, requiring an inclusive specification and careful communication of the moral ODD.

\subsubsection{Practical considerations} 

The specification and clear communication of the moral ODD support relevant humans (e.g., users, designers, developers) to be aware of the moral implications of the system's actions and their responsibility for these actions, thereby supporting the tracing condition of meaningful human control. Furthermore, if the operation of the AI agent remains confined within the boundaries of what it ``can do'' and ``ought to do'', the tracking condition of meaningful human control is supported as well, as this makes the human-AI system more responsive to human understanding of what is the morally appropriate domain and mode of operation. Achieving these benefits requires that: 
\begin{enumerate*}[label=(\arabic*)]
    \item the moral ODD be explicitly defined; 
    \item the AI agent embed concrete solutions to constrain the actions of the human-AI system within the boundaries of the ODD.
\end{enumerate*}

To define the moral ODD, designers and developers need to engage with fundamental questions of what are the elements composing the moral ODD and how do the features of each element affect the system's behavior. The process starts with an ontological modelling of the environment(s) in which the human-AI system is expected to operate. Such complex assemblage of elements and relationships could be meaningfully represented within the moral ODD by making use of principles from existing research on software applications where ontologies are developed to enable context-aware computing systems \cite{bettini_survey_2010,cabrera_3lconont_2019}. The mapping of material and relational elements characterizing a domain should be complemented with an investigation of what might be the morally relevant reasons, what they represent in the specific context, assumptions and consequences related to the system operation. Such understanding of the moral landscape of an AI agent under development could be built by means of extensive literature and case reviews \cite{coeckelbergh_drones_2013,galliott_military_2015,childress_public_2002}, participatory approaches such as interviews, interactive workshops, and value-oriented coding of qualitative responses \cite{friedman_value_2019}, which can be supported by natural language processing algorithms  \cite{liscio_axies_2021}.

How to satisfy the second requirement (constraining the AI agent to the boundaries of the moral ODD) varies according to the constituent elements of the moral ODD. When constraining the material and relational aspects of the system behavior, approaches developed in the automotive and aircraft domains can be a useful reference, e.g., risk-based path planning strategies for unmanned aircraft systems in populated areas \cite{primatesta_riskbased_2020} and geofencing  \cite{maiouak_dynamic_2019}. Relational aspects can be addressed through envelope protection. In the aircraft domain, flight envelope protection systems prevent the pilot from making control commands that drive the aircraft outside its operational boundaries, a concept that has also been adopted for unmanned aerial vehicles \cite{yavrucuk_envelope_2009}. This concept could be extended beyond the aircraft domain, and become a more general design pattern for constraining the relational elements of the moral ODD in the systems involving both embodied and non-embodied AI agents \cite{robbins_ai_2020}. 

Moral constraints are arguably the most challenging to enforce. One possible way of imposing them is to set probabilistic guarantees on system outcomes \cite{thomas_preventing_2019}. However, these approaches might not hold in real-world applications. Due to the non-quantifiable nature of morally relevant elements, as well as moral disagreements among humans, the boundaries of the moral ODD will remain blurred \cite{burton_mind_2020}. Hence, it is crucial that humans, not AI agents, are empowered to be aware of their responsibilities in order to make conscious decisions if and when the human-AI system should deviate from the boundaries defined by the moral ODD. The assessment of whether and how an AI agent is confined to the moral ODD is not a binary check, but rather a contextualized and deliberated analysis of the interaction between the AI agent, human agents, and the social, physical, ethical, and legal environment surrounding them. Humans, to conclude, should have an understanding of such blurry boundaries of the moral ODD and their responsibility to meaningfully control the AI agent in this process. Importantly, this includes the possibility of deciding that the use of an AI agent is not acceptable in certain contexts.

\subsection{Property 2. Human and AI agents have appropriate and mutually compatible representations of the human-AI system and its context}

For a human-AI system to perform its function, both humans and AI agents within the system should have some form of representations of the involved tasks, role distributions, desired outcomes, the environment, mutual capabilities and limitations. Such representations are often referred to as \textit{mental models}; these models enable agents to describe, explain and predict the behavior of the system and decide which actions to take \cite{johnson_coactive_2014,jonker_shared_2011,wilson_mental_1989}. 

\begin{figure}[!ht]
    \centering
    \includegraphics[width=\textwidth]{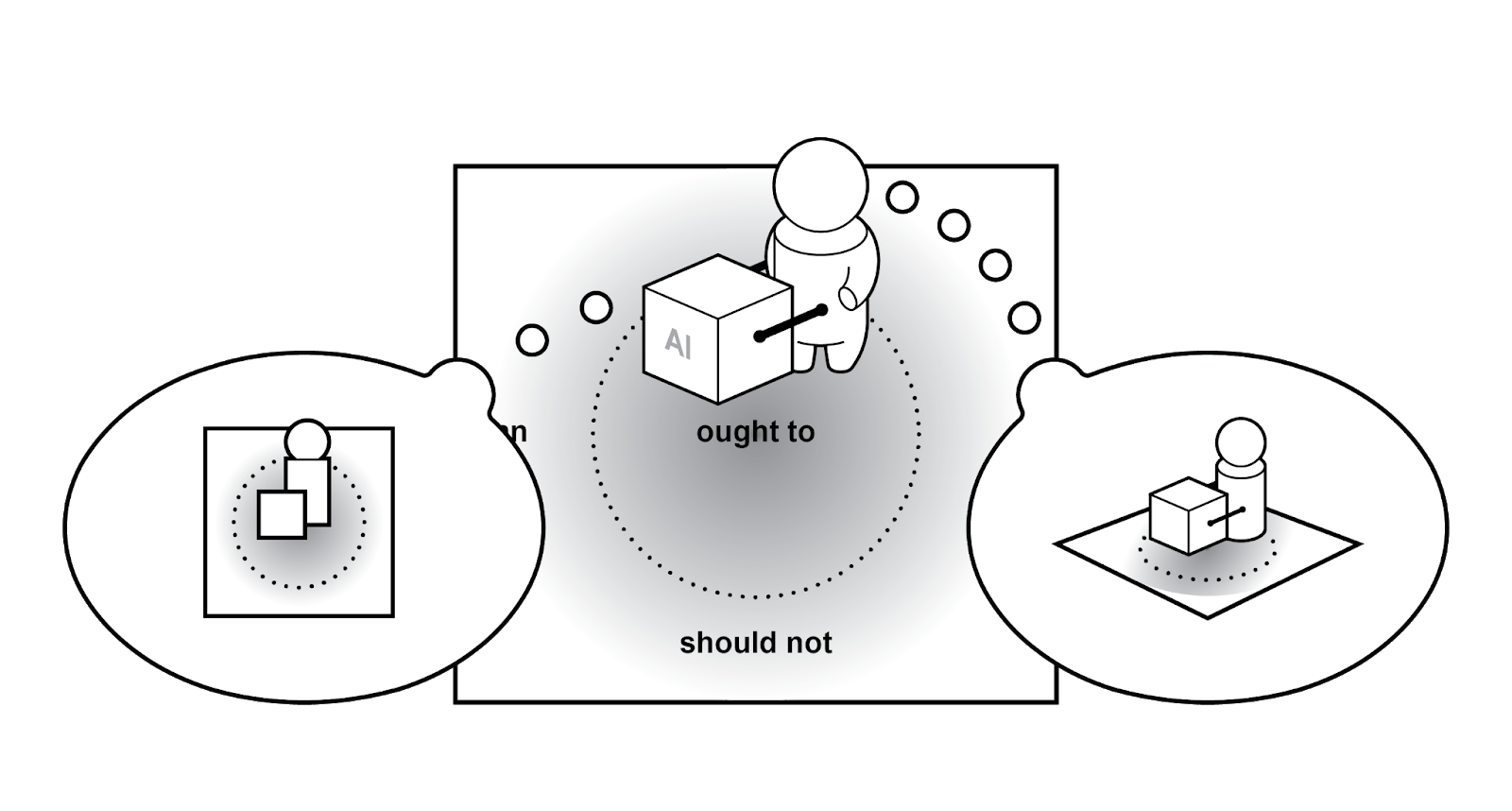}
    \caption{Property 2: The human and AI agents have appropriate and mutually compatible representations of the human-AI system and of each other’s abilities and boundaries.}
    \label{fig:prop2}
\end{figure}

Shared representations, i.e., representations that are mutually compatible between human and AI agents within the system, allow the agents to have appropriate understanding of each other, the task, and the environment \cite{jonker_shared_2011}, which facilitates agents to cooperate, adapt to changes, and respond to relevant human reasons. To ensure safe operation of the system, agents should also have a shared representation of each other's abilities and limitations. Specifically, the AI agents should account for humans' inherent physical and cognitive limitations, while human agents should account for the AI agents' limitations to avoid issues such as overreliance \cite{lee_trust_2004}. Furthermore --- crucial to achieve meaningful human control --- these shared representations should include the human reasons identified in the moral ODD (Figure \ref{fig:prop2}), which can change over time and across contexts. Due to the dynamic nature of elements of the shared representations, the human and AI agents should be able to update their representations of the potentially changing reasons accordingly. 

Incompatibility between representations could result in the lack of responsiveness to human reasons, thereby leading to undesired outcomes with significant moral consequences. For example, inconsistent mental models between a human driver and automated vehicle about ``who has the control authority'', in which the human driver believes that the automated vehicle has control and vice versa, could result in a critical and unsafe system state \cite{flemisch_dynamic_2012}.

\subsubsection{Practical considerations} 

In order for the agents' shared representations to facilitate the system's tracking of relevant human reasons, the system designers first need to define which aspects of the system and its context (including relevant humans, AI agents, the environment, and the moral ODD) each agent should have a representation of. The process of determining what kinds of representations are needed will be context-specific and depend on the moral ODD of the system. A useful approach to determine the necessary representations and to translate these high-level concepts into practical design requirements is co-active design \cite{johnson_coactive_2014}. Specific to building and maintaining shared representations, this approach provides guidelines on how to establish observability and predictability between the human and AI agents, including what needs to be communicated and when \cite{jonker_shared_2011}. 

Representations can include practical matters such as task allocation, role distribution and system limits, but also understanding of how humans perceive the AI agents, human acceptance of and trust in the human-AI system, humans values and social norms. This should also include determining the appropriate level of representation. For instance, for an automated vehicle to interact with a pedestrian, the designers need to determine whether it suffices for the vehicle to have a representation of just the location of a pedestrian on the road and their movement trajectory, or also the height and age of that human, their goals and intentions. In the context of AI-based hiring, a key aspect requiring shared representation is the meaning of competence. In particular, the meanings of soft skills, such as teamwork and creativity, are highly fluid, context-dependent, and contestable. Therefore, aligning the job-specific meaning of competence among job seekers, employers, and any AI agent involved in the hiring process is critical.

Once the representations required for each agent are defined, the design and engineering choices need to sufficiently take these into account. Specifically, such choices should facilitate 
\begin{enumerate*}[label=(\arabic*)]
    \item AI agents to build and maintain representations of the humans and their reasons, and 
    \item humans to form mental models of AI agents and the overall human-AI system.
\end{enumerate*}
These shared representations can be achieved through various combinations of implicit (e.g., through interaction between agents) or explicit ways (e.g., by means of human training, verbal communication). For example, to allow humans to build and maintain a representation of an AI agent, it can be developed to be observable and predictable implicitly through its design (e.g., glass-box design \cite{alertubella_governance_2019}), allowing the operator to better understand the AI agent's decision-making. Ecological interface design can also leverage knowledge on human information processing to design human-AI interfaces that are optimally suited to convey complex data in a comprehensible manner \cite{vicente_ecological_1992}. Maintaining accurate representations during the human-AI system's deployment can also occur through interaction, either implicitly (e.g., through intent inference from observed behavior) or explicitly (e.g., explicit verbal or written messages). For example, an AI system can probe through behavior whether the human is aware of it's intentions before committing to a decision~\cite{sadigh_planning_2018}.  

For the AI agents to have appropriate representations of human agents, the assumptions about human intentions and behavior adopted by AI agents (either implicitly or explicitly) need to be validated. This can be aided by incorporating theoretically grounded and empirically validated models of humans in the interaction-planning algorithms of AI agents~\cite{schurmann_personalizing_2020,siebinga_validating_2021}, or by augmenting bottom-up, machine-learned representations with top-down symbolic representations~\cite{vanbekkum_modular_2021,marcus_next_2020}. An alternative approach, value alignment \cite{gabriel2020artificial}, aims to mitigate the problems that arise when autonomous systems operate with inappropriate objectives. In particular, inverse reinforcement learning (IRL), which is often used in value alignment, aims to infer the ``reward function'' of an agent from observations of the agent's behavior, also in cooperative partial-information settings (cooperative IRL) \cite{hadfield-menell_cooperative_2016}. Although IRL is likely not sufficient to infer human preferences from observed behaviour since human planning systematically deviates from the assumed global rationality \cite{armstrong_occam_}, such approaches could still support agents to maintain aligned shared representations \cite{peysakhovich_reinforcement_2019}.

\subsection{Property 3. The relevant humans and AI agents have ability and authority to control the system so that humans can act upon their responsibility}

\begin{figure}[!ht]
    \centering
    \includegraphics[width=\textwidth]{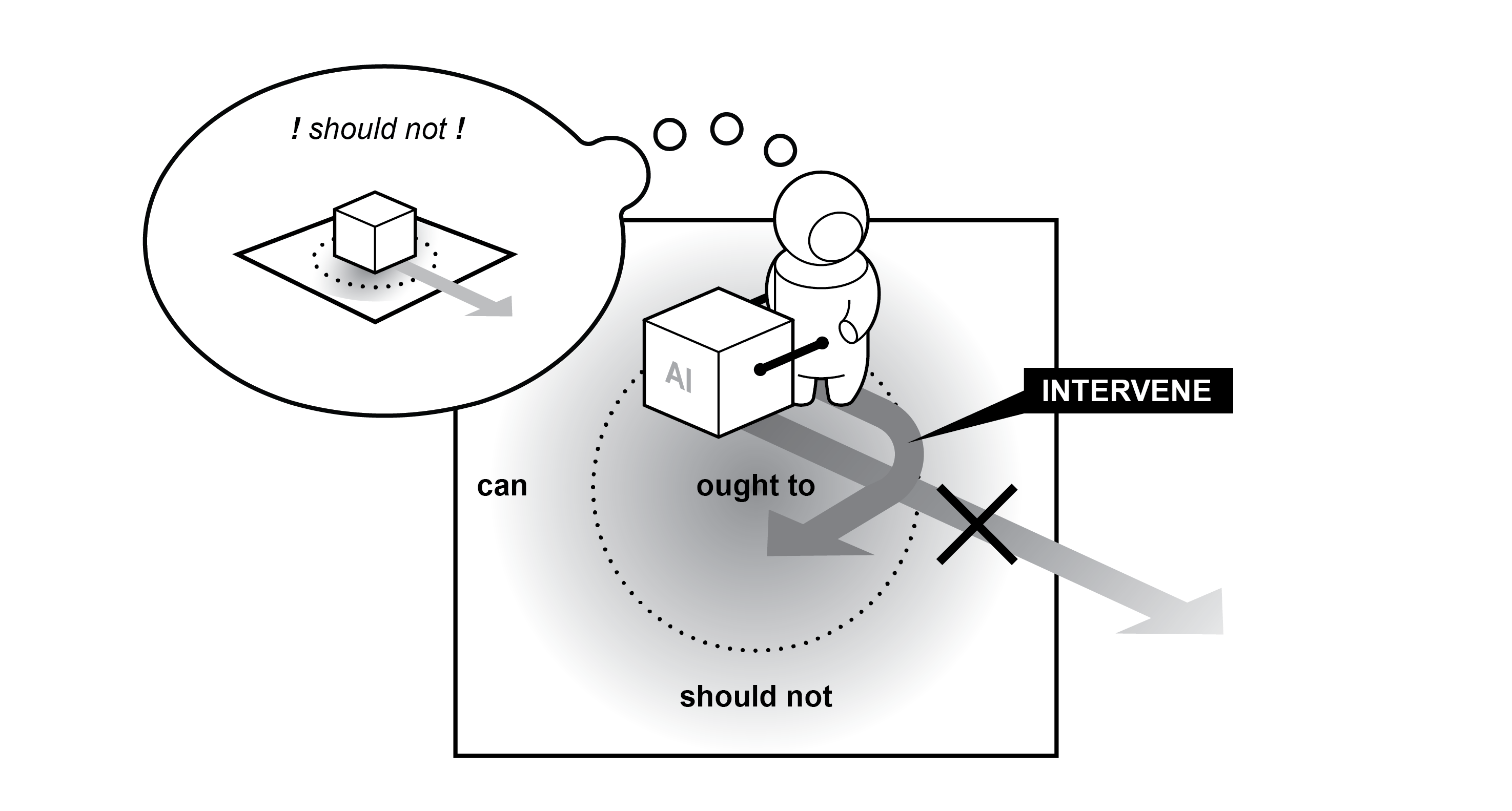}
    \caption{Property 3: The relevant humans and AI agents have the ability and authority to control the system so that humans can act upon their responsibility, e.g., if the human recognizes that a given situation might bring the system outside the moral ODD, they can intervene to avoid this.}
    \label{fig:prop3}
\end{figure}

Relevant humans should not be considered just mere subjects to be blamed in case something goes wrong, i.e., an ethical or legal scapegoat for situations when the system goes outside the moral ODD. They should rather be in a position to act upon their moral responsibility by influencing the AI system throughout its operation, and to bring the system back to the moral ODD if needed (Figure \ref{fig:prop3}). 

This is only possible when the distribution of roles and control authority between humans and AI (``who is doing what and who is in charge of what'') is consistent with their individual and combined abilities, including reasonable mechanism for overruling the AI agent through intervening and correcting behavior, setting new goals, or delegating sub-tasks. 



Flemisch et al. \cite{flemisch_dynamic_2012} provide a thorough account on the importance of an appropriate balance between an agent’s ability, authority, and responsibility in human-machine systems: ability to control should not be smaller than control authority, and control authority should not be smaller than responsibility. We argue that this account applies to complex human-AI systems as well. The \textit{ability} of a human or AI agent includes their skill and competence to perceive the state of a system and the environment. This also includes a way to acquire and analyze relevant information, to make a decision to act, and to perform that action appropriately \cite{parasuraman_model_2000}. Ability also includes the resources at their disposal, such as tools (an autonomous vehicle without a steering wheel would severely hamper the human's ability to control the vehicle's direction; job candidates’ ability to represent themselves would be heavily impaired by the lack of a feedback mechanism) or time (an automated vehicle that would wait until the very last second to alert the driver of a dangerous situation also limits the driver’s ability to direct the vehicle to safety; an employer would have no control and understanding of an AI-based hiring system if assessment of candidates would be provided only after the selection process finishes).

The understanding of an (AI or human) agent's \textit{ability} is intrinsically related to the socio-technical context in which the system is embedded. Hence, it is important that tasks are distributed according to the agent's ability in the context, not only from a functional perspective but also accounting for the values and norms intrinsic to the activity. Approaches such as the nature-of-activities \cite{santonidesio_who_2014,santonidesio_when_2016}, under the umbrella of Value Sensitive Design \cite{friedman_value_2019}, can support the understanding of which set of tasks should be (partially or totally) delegated or shared with AI agents, and which should be left exclusively to humans. Given the collaborative nature of many human-AI systems, team design patterns can be used as an intuitive graphical language for describing and communicating to the team the design choices that influence how humans and AI agents collaborate \cite{vandiggelen_team_2019,vanderwaa_allocation_2020}.

The second component of the account proposed in \cite{flemisch_dynamic_2012} is control \textit{authority}, i.e., the degree to which a human or AI agent is enabled to execute control. 
Consistency between authority and ability requires that an agent’s authority does not exceed their ability. And similarly, responsibility should not exceed authority. Thus, an agent should be responsible only for tasks they have authority to perform, and they should have authority only over tasks they are able to perform. A key implication of this consistency is that control is exerted by the agent that has sufficient ability and authority, and more responsibility is carried by the agents that exert more control. While ability and authority are attributes that both human and AI agents possess, we consider responsibility as a human-only quality. Therefore, the ability and authority of a human-AI system must be traced to responsibilities of relevant humans, e.g., engineers, designers, operators, users, and managers.

In the automated vehicle case, the driver has authority to control the vehicle by accelerating, breaking and steering, as well to take over control authority at any time. In the case of AI-based hiring, employers' authority includes setting a threshold for a passing score and deciding who to hire. Simply giving human agents final authority by design, without ensuring proper ability, is not sufficient to empower humans to act upon their moral responsibility. For example, a driver may have final authority over a fully autonomous car, but the driver's loss of situational awareness, or even skill degradation as a result of systematic lack of engagement in the driving task, will limit the driver's ability to exert that control authority \cite{flemisch_dynamic_2012,heikoop_human_2019,kyriakidis_human_2019}. The same might happen for a manager with final authority over who to hire, 

if they merely sign off on the hiring recommendations of the AI agent, without substantively engaging in the assessment process themselves.

\subsubsection{Practical considerations} 

As \textit{authority} should not be smaller than \textit{ability}, it is important to build a baseline understanding of the abilities of human and AI agents and evaluate their consistency with the control authority provided by the system's design. From the human side, human factors literature \cite{salvendy_handbook_2012} can support the identification of a realistic baseline on human ability by applying psychological and physiological principles to understand challenges that are likely to arise in human-AI interaction~\cite{sujan_human_2019,kyriakidis_human_2019}. From the AI side, a proper understanding of ability should not only be task-oriented (e.g., measuring performance from data sets against benchmark), but also behavior-oriented. Approaches to understand AI ability in context include approaches inspired by human cognitive tests, information theory \cite{hernandez-orallojose_evaluation_2017}, and ethology (related to animal behavior) \cite{rahwan_machine_2019}. Designing for appropriate authority and ability also requires us to expand the scope of design from human-AI interactions to social and organizational practices \cite{santonidesio_meaningful_2018}. Human training, oversight procedures, administrative discretion, and policy are just a few examples of organizational elements that significantly determine and shape agents' authority and ability.

Design, training and technological development may ``expand'' or ``shrink'' agents' abilities through innovation, including training humans for new skills and equipping AI agents with new technological capabilities, or achieving more through interaction between humans and AI and their combined abilities. From the AI side, especially for machine learning-based systems, as the relation between the input data and the target variable changes over time, concept drift methodologies can be applied to identify new situations which might impact the AI agent's ability to respond to new situations  \cite{gama_survey_2014,lu_learning_2018}. From the human side, interaction with technology might lead to behavioral adaptation and unwanted situations e.g., speeding when driving with intelligent steering assistance provided by an automated vehicle~\cite{melman_does_2017}, decreasing human's ability to keep the system within the moral ODD. 
In such situations, the human-AI system might move to a fallback state \cite{christian_partially_2013} or attract the driver’s attention back to the supervision task thus restoring the driver’s ability to act upon their ultimate responsibility for the vehicle’s operation.

Shared control is a promising approach to keep a balance between control ability and authority, with relevant applications in the domain of automated vehicles, robot-assisted surgery, brain-machine interfaces, and learning \cite{abbink_topology_2018}. In shared control, the human(s) and the AI agents(s) are interacting congruently in a perception-action cycle to perform a dynamic task, i.e., control authority is not attributed either to the human or to the AI agent, but is shared among them \cite{abbink_haptic_2012}. Shared control could be particularly useful in human-AI systems that need to act in complex situations that can rapidly change beyond the envisioned moral ODD, and where rapid human adaptation and intervention is needed. 

\subsection{Property 4. Actions of the AI agents are explicitly linked to actions of humans who are aware of their moral responsibility}

Satisfying the first three properties ensures that relevant humans are capable of acting upon their moral responsibility (property 3), are aware of the moral implications of the system's actions (property 1), and have shared representations with AI agents (property 2). Yet, what is left undiscussed is the requirement to ensure that the effects of the system’s actions are traceable to the relevant humans’ moral understanding. 

To trace any consequence of the human-AI system’s operation to a proper moral understanding of relevant humans, there should be explicit, explainable and inspectable link(s) between actions of the system and corresponding human morally-loaded decisions and actions. We acknowledge that such link(s) might be a more demanding form of tracing than what was originally proposed in \cite{santonidesio_meaningful_2018}, nevertheless we deem it necessary to enable the tracing condition to be inspectable. Furthermore, we argue that moral understanding of the system’s effects should be demonstrated by, at least, those humans who make decisions with moral implications on the design, deployment, or use of the system, even if the actions that bring a human decision to life are executed by the AI agent. Hence, all relevant human decisions related to e.g., design, use, policy must be explicitly logged and reported \cite{alertubella_governance_2019}, in order to link actions of the AI agents to relevant decisions, preferences, or actions of humans who are aware of the system’s possible effects in the world.

\begin{figure}[!ht]
    \centering
    \includegraphics[width=\textwidth]{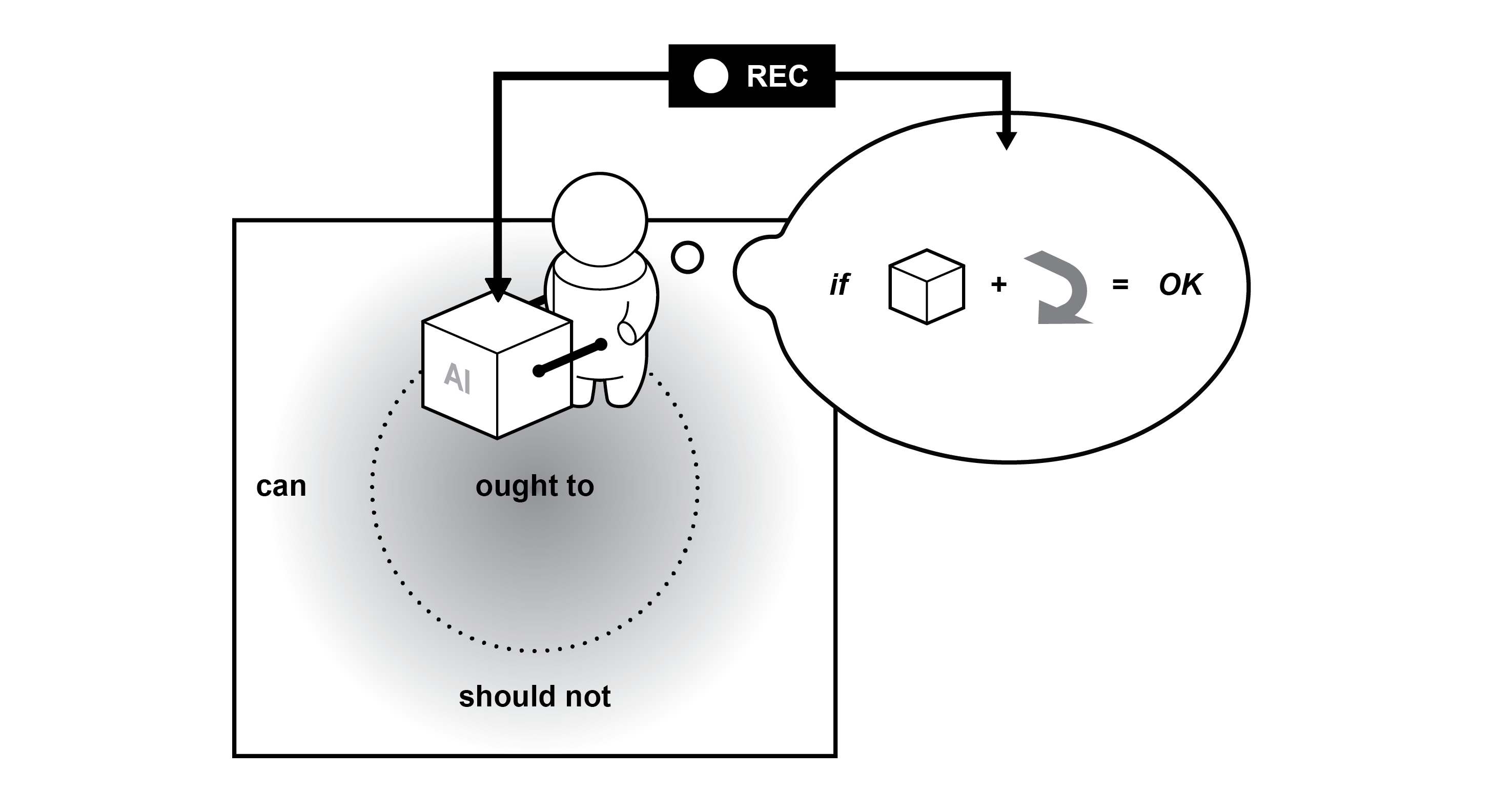}
    \caption{Property 4: Actions of the AI agents are explicitly linked to actions of humans who are aware of their moral responsibility. The forward link starts on the human and indicates that whenever a human makes a decision with moral implications that affects the system's behavior, that human should be aware of their moral responsibility. The backward link looks at actions of the human-AI system and links it to previous human decisions (e.g., designers, users).}
    \label{fig:prop4}
\end{figure}

Even if all relevant humans made their decisions responsibly and with full awareness of their possible moral implications, the lack of a readily identifiable link from a given action of the AI agent to the underlying human decisions would still result in loss of tracing. The links between actions of the systems and corresponding human morally-loaded decisions and actions need to be explicitly identifiable in two ways (Figure \ref{fig:prop4}):
\begin{enumerate}[label=(\arabic*)]
    \item \textit{Forward link}: whenever a human within the human-AI system makes a decision with moral implications (e.g., on the design, deployment, or use of the system), that human should be aware of their moral responsibility associated with that decision, even if the actions that bring this decision to life are executed by the AI agent.
    \item \textit{Backward link}: for any consequence of the actions of the human-AI system, the human decisions and actions leading to that outcome should be readily identifiable.
\end{enumerate}

\subsubsection{Practical considerations} 

Enabling the \textit{forward link} from human moral understanding to AI agents’ actions relates to the epistemic condition (also called knowledge condition) of moral responsibility, which posits that humans should be aware of their responsibility at the time of a decision \cite{aristotle_nicomachean_1999,coeckelbergh_ai_2020}. Hence, the human-AI system should be designed in a way that simplifies and aids achieving moral awareness. This requires explicit links between design choices and stakeholder interpretations of moral reasons that are at stake. Values hierarchies \cite{vandepoel_translating_2013} provide a structured and transparent approach to map relations between design choices and normative requirements. A value hierarchy visualizes the gradual specification of broad moral notions, such as moral responsibility, into context-dependent properties or capabilities the system should exhibit, and further into concrete socio-technical design requirements. Such a structured mapping can equip stakeholders with the means to deliberate design choices in a manner that explicitly links each choice to relevant aspects of moral responsibility. These deliberations, as well as the accompanying rich body of empirical and conceptual research must be well documented, inspectable, and legible. This kind of transparency also supports the \textit{backward link} between the system's actions and the design choices made by relevant humans.

Furthermore, requirements such as explainability of the system’s actions can be essential in effectively empowering human moral awareness. Since its early works, the field of explainable AI has increased its scope from explaining complex models to technical experts towards placing the target audience as a key aspect \cite{barredoarrieta_explainable_2020}. Given a certain human or group of humans as target audience, we see explainability in the context of supporting the forward link as clearly presenting the link between the system's actions and human moral awareness, as well as their alignment to the moral ODD. For example, consider an automated vehicle which slows down and pulls off the road after it recognizes a car accident \cite{mecacci_meaningful_2020}. Right after that the vehicle should then remind the driver of their duty to provide assistance to possible victims in the accident. In the context of AI-based hiring, explanations of assessment scores in language that directly links observed job seeker performances to job-specific meanings of competence can help employers, job seekers, designers, and developers better outline the boundaries of moral ODD during design phase. For example, this can help reveal whether there is misalignment between conceptions of competence among the human agents and the AI algorithm.

In complex socio-technical environments the establishment of links between human moral awareness and actions of a human-AI system is complicated by the ``problem of many hands'', which happens when more than one agent contributes to a decision. It becomes less clear who is morally and legally responsible for its consequences \cite{vandepoel_ethics_2011}. The ``problem of many things'' complicates this further: there are not only many (human) hands, but also many different technologies interacting and influencing each others, be it multiple AI agents or the interplay between sensors, processing units, and actuators \cite{coeckelbergh_artificial_2019}.
In case of unintended consequences of the AI agent’s actions, this complexity can hinder the backward link, i.e., tracing the responsibility back to individual human decisions. This challenge calls for systemic, socio-technical design interventions that jointly consider social infrastructure (e.g., organizational processes, policy), physical infrastructure, and the AI agents that are part of these infrastructures. 

Recent developments using information theory to quantify human causal responsibility \cite{douer_responsibility_2020} can provide relevant insight for the design and development of appropriate forward and backward links, by providing a model with which hypotheses can be tested. However, simplifying assumptions used in this research need to be addressed to account for more realistic settings. Methods from social sciences, e.g., Actor-Network Theory (ANT) \cite{latour_reassembling_2005} can support the development of tracing networks of association amongst many actors, which can help understand how, for example, humans may offload value-laden behavior onto the technology around us. In the ``sociology of a door closer'', \cite{johnson_mixing_1988} describes how we made door closers the element in the assembly that manifests politeness by ensuring the door closes softly and gradually, even as the human actors may barge through without any action to regulate the door). This sort of division of moral-labor should not be done mindlessly, it requires human decisions to be analyzed and their relation to the moral ODD to be carefully analyzed. 

Although establishing explicit links between human decisions, human moral awareness, and actions of the AI agents is challenging, they allow appropriate post-hoc attribution of backward-looking responsibility for unintended consequences, helping to avoid responsibility gaps and prevent similar events from repeating in the future. It also facilitates forward-looking responsibility by creating an incentive for the relevant humans to proactively reflect on the consequences of their decisions (design choices, operational control, interactions, etc.).

\subsection{Summary of the four properties}
We summarize the proposed properties of systems under meaningful human control as follows:
\begin{itemize}[]
    \item \textit{Property 1:} The human-AI system has an explicit moral operational design domain (moral ODD) and the AI agent adheres to the boundaries of this domain. 
    \item \textit{Property 2:} Human and AI agents have appropriate and mutually compatible representations of the human-AI system and its context.
    \item \textit{Property 3:} The relevant agents have ability and authority to control the system so that humans can act upon their responsibility.
    \item \textit{Property 4:} Actions of the AI agents are explicitly linked to actions of humans who are aware of their moral responsibility.
\end{itemize}

In our view, these properties are constructive as well as open: they can serve as practical tools for supporting the design, development and evaluation of human-AI systems, while being applicable to diverse types of systems (as illustrated by the cases of automated vehicles and AI-based hiring).

Although the properties are not sufficient for a system to be under meaningful human control, we deem them necessary from a design perspective: while a system developed to possess all these properties may still not be fully under meaningful human control, we believe that completely missing one of these properties would imply that the human-AI system is not under meaningful human control. Moreover, each property is non-binary and necessarily multidimensional.  Consequently, improving the system to some extent according to one or more of the four properties will lead to better tracking or tracing, and therefore, more meaningful human control over that system. That said, defining “how much of each property is sufficient" in a given context would generally require a thorough qualitative and situated analysis.

Furthermore, these four properties in themselves do not immediately translate to concrete design guidelines; metrics, algorithms, and methodologies needed to implement the properties are context- and system-specific. Yet, the properties provide explicit anchors for connecting to existing frameworks and methodologies across the design and engineering domains.

\section{The broader picture}\label{sec_disc}

In addition to establishing explicit links between the concept of meaningful human control and existing frameworks across the design and engineering domains, the four proposed properties unveil a range of new methodological questions and challenges on the path to practically implementing systems under meaningful human control.

\textit{Designing for meaningful human control requires designing for emergence.} We argue that improving the human-AI system according to the properties we presented will lead to better tracking and tracing, and therefore more meaningful human control over that system. However, that does not provide an answer to the critical question: how much meaningful human control is sufficient in a given context? We believe these uncharted waters need to be explored through practice-based research that aims to responsibly develop human-AI systems, while ensuring inclusive and transparent collaborations among stakeholders and safe and rigorous evaluation of concepts and designs. On the one hand, it is reasonable to expect that socio-technical design requirements that act for the sake of meaningful human control properties will vary across societal and application domains. On the other hand, given a sufficient level of conceptual abstraction, a common basic set of system properties that will prove practically helpful and robust across different societal domains can inform both bottom-up practice and top-down regulation towards meaningful human control. However, design and regulation cannot account for every detail of a system’s processes, interactions, components in a deterministic, top-down fashion. In fact, the socio-technical complexity of human-AI systems and the inherent uncertainty of some aspects of their operation call for designing for emergence \cite{pendleton-jullian_design_2018}, where the focus shifts to designing the social, physical, and technical infrastructures that jointly provide favorable conditions for interactions between agents to lead to emergence of desirable system properties and behaviors.

\textit{Meaningful human control is necessary but not sufficient for ethical AI.} Meaningful human control over AI relates to the broader scope of AI ethics in the sense that designing for meaningful human control means designing for human moral responsibility. That is a critical aspect of ethical design of human-AI systems, but by itself it is not sufficient to ensure other crucial aspects of ethical design and operation, such as protection of human rights and environmental sustainability. In fact, it is possible for a human-AI system to be under meaningful human control with respect to some relevant humans, yet result in outcomes that are considered morally unacceptable by society at large \cite{santonidesio_meaningful_2018}. Meaningful human control ensures that humans are aware of and are equipped to act upon their responsibility, and that the human-AI system is responsive to human moral reasons. But it does not prevent humans from consciously designing and operating the human-AI system in an unethical way. Therefore, meaningful human control must be part of a larger set of design objectives that collectively align the human-AI system with societal values and norms. 

\textit{Transdisciplinary practices are vital to achieve meaningful human control over AI.} One of the most prominent challenges threaded throughout the four properties may also be the most rewarding opportunity: the inherent need for a socio-technical design process that crosses disciplinary boundaries. Each of the four properties and meaningful human control as a whole is an endeavor that is not solvable by a single discipline. It is a systemic, socio-technical puzzle in which computer scientists, designers, engineers, social scientists, legal practitioners, and crucially, the societal stakeholders in question, each hold an essential piece of the puzzle. Hence, the only way to ``walk the walk'' is to move forward together, forming a transdisciplinary practice based on continuous mutual learning \cite{vanderbijl-brouwer_systemic_2020} among both academic and non-academic stakeholders. While this is undoubtedly a challenge, it may prove to be a rewarding opportunity for socially inclusive innovation that puts human moral responsibility front and center.

\section{Conclusion}\label{sec_concl}

In this article, we address the issue of responsibility gaps in design and use of AI systems, and argue in favor of the concept of meaningful human control as a principle to mitigate them. To the current discourse surrounding meaningful human control, we contribute with a set of four actionable system properties and related approaches useful for implementing them in practice. These properties unpack the tracking and tracing conditions of meaningful human control \cite{santonidesio_meaningful_2018} and provide a significant step forward toward its operationalization. Even though these properties may not be sufficient to completely ensure meaningful human control for all possible situations, we deem them necessary, and as such they help translate the tracking and tracing conditions into more tangible and designable requirements for human-AI systems. Our properties build upon and expand existing conceptual frameworks and methodologies across the design and engineering domains, such as the notion of operational design domain \cite{czarnecki_operational_2018}, ontological modeling \cite{cabrera_3lconont_2019}, co-active design \cite{johnson_coactive_2014}, shared mental models \cite{jonker_shared_2011}, shared control \cite{abbink_topology_2018}, value alignment \cite{gabriel2020artificial}, and consistency of ability, control authority, and responsibility \cite{flemisch_dynamic_2012}. With these four properties we have realized two goals: \begin{enumerate*}[label=(\arabic*)]
    \item contributed to closing the gap between the philosophical theory and practice of designing systems under meaningful human control, and
    \item explicitly link meaningful human control to existing frameworks and methodologies across disciplines that can support design and development of human-AI systems.
\end{enumerate*} 


Societal impacts and the issue of responsibility gaps in the use of AI today puts forward meaningful human control as one of the central concepts when discussing trustworthy and responsible AI, and we think it should also take central place on AI development. We believe this work will enable researchers and practitioners to take actionable steps towards the design and development of systems under meaningful human control, enabling many of the promised benefits of AI while maintaining human responsibility and control.

\section*{Competing interest}
The authors have no competing interests to declare that are relevant to the content of this article.

\section*{Acknowledgement}
We thank Filippo Santoni de Sio for helpful comments on the earlier version of this manuscript.

\bibliography{sn-bibliography}


\end{document}